\documentclass[cameraready]{Interspeech}

\usepackage{amsmath,amssymb,amsthm}
\usepackage{amsfonts} 
\usepackage{gensymb}  
\usepackage{multicol}
\usepackage{multirow}
\usepackage{pifont}
\usepackage{xpatch,xcolor}
\usepackage{hyperref}
\usepackage{caption}
\usepackage{subcaption}
\usepackage{bbm}
\usepackage{booktabs}
\usepackage{makecell}

\title{Joint Learning of Covariance Estimation and White Noise Gain \\ for Robust MVDR Beamforming
}

\author[affiliation={1}]{Yongyi}{Deng}
\author[affiliation={1}]{Hanchen}{Pei}
\author[affiliation={2}]{Jianbo}{Ma}
\author[affiliation={1}, correspondingauthor]{Gongping}{Huang}
\author[affiliation={3}]{Jingdong}{Chen}
\author[affiliation={4}]{Jacob}{Benesty}

\address{
$^1$ School of Electronic Information, Wuhan University, Wuhan, Hubei, China \\
$^2$ Dolby Laboratories \\
$^3$ CIAIC, Northwestern Polytechnical University, Xi'an, Shaanxi, China \\
$^4$ INRS-EMT, University of Quebec, Montreal, QC, Canada
}

\email{dyy520@whu.edu.cn}


\usepackage{comment}

\begin{document}

\maketitle

\begin{abstract}
The minimum variance distortionless response (MVDR) beamformer is widely used for multichannel speech enhancement due to strong noise suppression while preserving target signals. In practice, its performance is sensitive to microphone self-noise and array mismatches. Existing approaches typically rely on fixed, manually tuned WNG thresholds or diagonal loading, leading to suboptimal performance under unknown or time-varying acoustic conditions. This paper proposes a data-driven MVDR framework that adaptively estimates the WNG constraint using a deep neural network. The network jointly predicts a time–frequency noise mask for covariance estimation and a frequency-dependent WNG threshold, enabling dynamic robustness–directivity control. A differentiable robust MVDR layer is integrated into the framework, allowing end-to-end optimization. Experiments demonstrate consistent improvements in speech quality and intelligibility over conventional fixed-WNG MVDR methods.
\end{abstract}

\section{Introduction}
\label{Sect-Intro}

Microphone array beamforming is a core technique in multichannel speech processing, with applications including voice capture, spatial audio recording, and environmental perception~\cite{brandstein2001microphone, benesty2018fundamentals, luo2024design, huang2022kronecker, cohen2025explainable}. Among existing approaches, the minimum variance distortionless response (MVDR) beamformer is particularly attractive~\cite{yamaoka2021time, kealey2024Unspuervised, tao2024learning, zhao2025diffusion}. The MVDR beamformer relies on accurate estimation of the spatial covariance matrix of the received signals~\cite{van1988beamforming, tavakoli2015pseudo, habets2010speech, huang2019new, moore2022compact, alam2015regularized}, which is commonly obtained using time-averaged statistics or voice activity detection (VAD)-based methods~\cite{moore2022compact, habets2010speech, heb2025microphone}. More recently, deep neural networks have been widely adopted to estimate time--frequency noise masks, enabling improved covariance matrix estimation and consequently enhanced MVDR performance~\cite{erdogan2016improved, heymann2016neural, higuchi2017online, higuchi2016robust}.
Despite its effectiveness, the MVDR beamformer is inherently sensitive to array imperfections and modeling errors, such as microphone self-noise, gain and phase mismatches, and sensor position inaccuracies. This sensitivity is commonly quantified by the white noise gain (WNG)~\cite{pei2025data}, which characterizes the robustness of a beamformer against spatially white noise and array uncertainties~\cite{cox1987robust, benesty2008microphone}. A low WNG indicates a high susceptibility to array mismatches and often leads to significant performance degradation in practical scenarios~\cite{lobato2020worst, ehrenberg2010sensitivity}. To improve robustness, WNG-constrained MVDR formulations have been proposed, where a minimum WNG level is enforced through diagonal loading or equivalent regularization strategies~\cite{chen2021robustness, benesty2023microphone, harmanci2000relationships, huang2022fundamental}.

In most existing robust MVDR approaches, the WNG threshold is fixed and selected empirically~\cite{benesty2008microphone, pan2013performance}. Such a fixed robustness setting is inherently suboptimal, as the optimal trade-off between robustness and spatial selectivity depends on the acoustic scene, noise characteristics, source configuration, and array geometry. Moreover, microphone arrays deployed in real-world devices often exhibit non-negligible variability due to manufacturing tolerances, aging effects, and device-specific self-noise levels, making it difficult to define a universal WNG threshold that generalizes well across different arrays and operating conditions. Consequently, robustness control is typically treated as a heuristic design choice rather than a signal-dependent property of the beamformer.
Recent learning-based beamforming methods have primarily focused on improving covariance matrix estimation, while the robustness control mechanism itself has received comparatively less attention. In particular, the WNG or diagonal loading factor is often regarded as a fixed hyperparameter or adjusted independently of the beamforming objective, without being explicitly integrated into an end-to-end optimization framework. As a result, the robustness--directivity trade-off remains externally imposed and cannot adapt to varying acoustic conditions or array mismatches.

To address these limitations, this work proposes a data-driven MVDR beamforming framework that learns robustness control directly from multichannel observations. Instead of treating the WNG as a manually tuned hyperparameter, it is interpreted as a latent physical control variable governing the robustness--directivity trade-off of the beamformer. A dual-branch neural network architecture is introduced to jointly estimate a time--frequency noise mask for covariance matrix estimation and a frequency-dependent WNG constraint for robust MVDR design. By embedding a differentiable WNG-constrained MVDR layer into the learning pipeline, the proposed framework enables optimization of both spatial statistics estimation and robustness control, without requiring explicit supervision on the WNG values. Experimental results demonstrate that the proposed method consistently outperforms conventional MVDR beamformers with fixed WNG thresholds, particularly under array mismatch conditions. 

\section{Signal Model and Problem Formulation}
\label{Sect-SM-PF}

Consider a microphone array consisting of $M$ sensors in a acoustic environment , capturing a desired source propagating from direction $\theta_{\mathrm{s}}$, the observation signal vector of length $M$ in the short-time Fourier transform (STFT) domain can be written as
\begin{align}
\label{Y-vect}
\mathbf{y}(k) 
&= \left[ \begin{array}{cccc}
	Y_1(n,k) & Y_2(n,k) & \cdots & Y_M(n,k) \end{array} \right]^T \nonumber \\
&= \mathbf{d}_{\theta_{\mathrm{s}}} (k) X\left( n,k \right) + \mathbf{v}\left( n,k \right),
\end{align}
where $Y_m(n,k)$ is the $m$th ($m=1,2,\ldots,M$) microphone signal of the array at time frame $n$ and frequency bin $k$, 
$\mathbf{d}_{\theta_{\mathrm{s}} }(k)$ is the signal propagation vector, the superscript $^T$ is the transpose operator, and $\mathbf{v}(n,k)$ is the noise signal vector defined similarly to $\mathbf{y}(n,k)$.In the assumed far-field setting, $\mathbf{d}_{\theta_{\mathrm{s}} }(k)$ is computed from the array geometry and the target direction. The desired signal and the noise are incoherent.

The covariance matrix of $\mathbf{y}\left( n,k \right)$ is given by
\begin{align}
	\mathbf{\Phi}_{\mathbf{y}}(k) 
	&= E \left[ \mathbf{y}(n,k) \mathbf{y}^H(n,k) \right] \nonumber \\
	&= \phi_X(k) \mathbf{d}_{\theta_{\mathrm{s}}} (k) \mathbf{d}_{\theta_{\mathrm{s}} }^H(k) + \mathbf{\Phi}_{\mathbf{v}} (k) \nonumber \\
	&= \phi_X(k) \mathbf{d}_{\theta_{\mathrm{s}}} (k) \mathbf{d}_{\theta_{\mathrm{s}} }^H(k) + \phi_V(k) \mathbf{\Gamma}_{\mathbf{v}} (k),
\end{align}
where $(\cdot)^H$ denotes the conjugate transpose,
$\phi_X(k) = E[|X(k)|^2]$ is the variance of $X(k)$,
$\mathbf{\Phi}_{\mathbf{v}}(k) = E[\mathbf{v}(k)\mathbf{v}^H(k)]$ is the spatial covariance matrix of $\mathbf{v}(k)$,
$E[\cdot]$ denotes the mathematical expectation, and
$\mathbf{\Gamma}_{\mathbf{v}}(k) = \mathbf{\Phi}_{\mathbf{v}}(k) / \phi_V(k)$ is the normalized spatial covariance matrix of the noise.

To extract the target signal from the multi-channel observation, a linear spatial filter $\mathbf{h}(k) \in \mathbb{C}^M$ is applied to the observation vector $\mathbf{y}(k)$.
In order to avoid signal distortion in the target direction, the beamformer weights are designed to satisfy the distortionless constraint:
\begin{align}
\label{const-oo}
\mathbf{h}^H(k) \mathbf{d}_{\theta_{\mathrm{s}}}(k) =1.
\end{align}
The SNR gain with weight vector $\mathbf{h}(k)$ can be expressed as
\begin{align}
	\label{SNR-gain-define}
	{\cal G}\left[ \mathbf{h} (k) \right]
		&= \frac{\left| \mathbf{h}^H(k) \mathbf{d}_{\theta_{\mathrm{s}}}(k) \right|^2}
	{\mathbf{h}^H(k) \, \mathbf{\Gamma}_{\mathbf{v}}(k) \, \mathbf{h}(k)},
\end{align}
which measures the improvement in SNR.
The MVDR beamformer is designed to maximize the SNR gain while ensuring the distortionless constraint, yields 
\begin{eqnarray}
	\label{SB-filt}
	\mathbf{h}_{\mathrm{MVDR}}(k) = 
	\frac{ \mathbf{\Gamma}_{\mathbf{v}}^{-1} (k) \mathbf{d}_{\theta_{\mathrm{s}}}(k) }
	{\mathbf{d}^H_{\theta_{\mathrm{s}}}(k) \mathbf{\Gamma}_{\mathbf{v}}^{-1} (k) 
		\mathbf{d}_{\theta_{\mathrm{s}}}(k) }.
\end{eqnarray}

The performance of a beamformer is highly susceptible to various array imperfections, such as microphone mismatches, position errors, and uncorrelated noise. To quantify its robustness against such disturbances, the WNG serves as a key metric:
\begin{align}
	\label{WNG-define}
	{\cal W}\left[ \mathbf{h} (k) \right] 
	&= \frac{\left| \mathbf{h}^H(k) \mathbf{d}_{\theta_{\mathrm{s}}}(k) \right|^2}
	{\mathbf{h}^H(k) \mathbf{h}(k)} .
\end{align}
For the MVDR beamformer, its WNG can become negative on the decibel scale, a phenomenon often referred to as white noise amplification. This implies that the beamformer is susceptible to array imperfections such as microphone position errors, gain mismatches, and phase offsets~\cite{elko2008microphone, huang2022fundamental}. 

In practice, array uncertainties are typically bounded within a certain range. This allows the use of a predefined WNG threshold, denoted by $\mathcal{W}_0$, to specify the minimum acceptable robustness level for the system, i.e., $\mathcal{W}(\mathbf{h}(k)) \ge \mathcal{W}_0$.
When designing such robust MVDR beamformers, two common strategies are typically adopted to determine an appropriate WNG level $\mathcal{W}_0$.
\begin{itemize}
\item
Constraining the WNG is equivalent to applying diagonal loading to the noise covariance matrix, i.e.,
$\mathbf{\Gamma}_{\mathbf{v},\epsilon}(k)=\mathbf{\Gamma}_{\mathbf{v}}(k)+\epsilon\mathbf{I}_M$, where $\epsilon>0$ denotes the diagonal loading factor.
Selecting a suitable WNG threshold is therefore equivalent to determining an appropriate loading parameter.
\item
Alternatively, the WNG-constrained MVDR problem can be formulated as a quadratic eigenvalue problem (QEP), which admits a closed-form solution for the beamformer weights.
This approach is computationally more efficient and avoids iterative parameter tuning~\cite{benesty2023microphone}.
\end{itemize}

\begin{figure*}[htbp]
  \centering
  \includegraphics[width=0.87\textwidth]{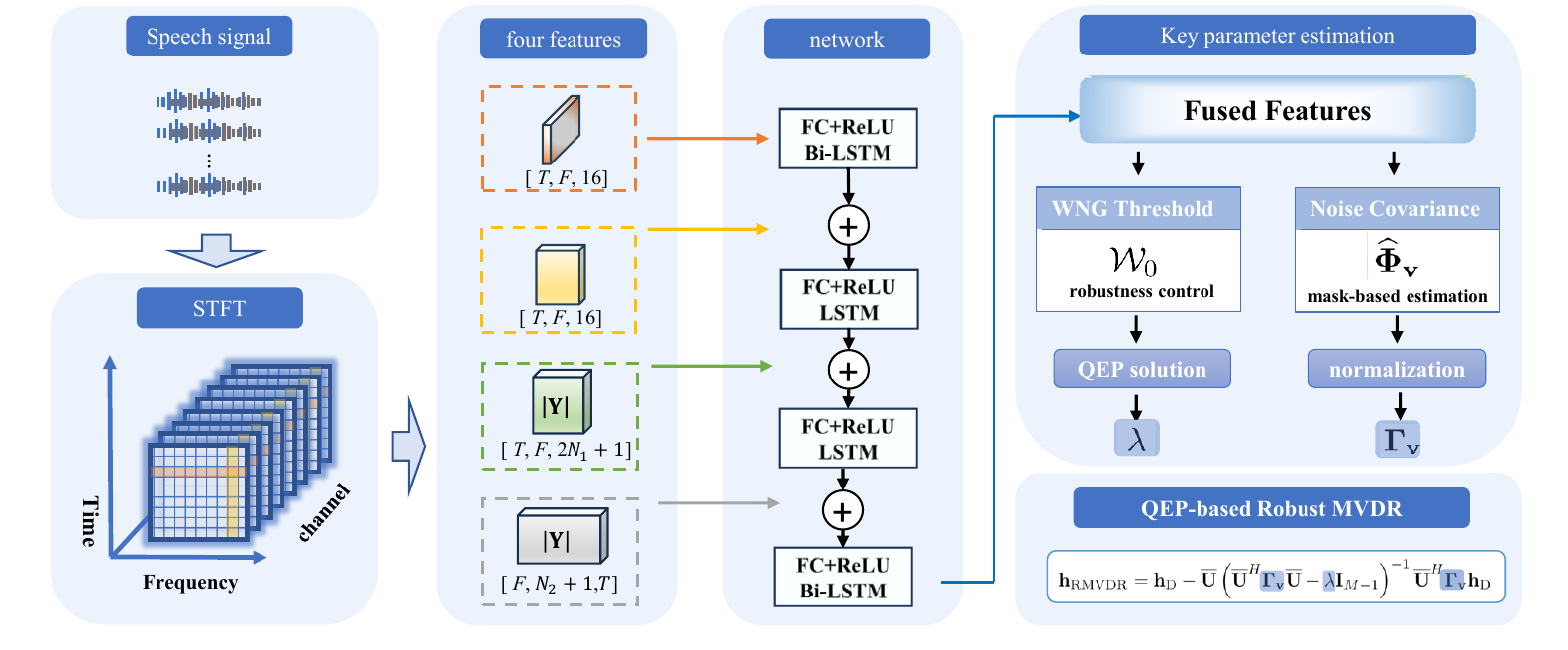}
  \caption{Overview of the proposed dual-branch network architecture for joint mask estimation and data-driven WNG prediction.}
  \label{fig:nn}
\end{figure*}

Following the theory in~\cite{benesty2023microphone}, any distortionless beamformer can be decomposed into the sum of two orthogonal components:
\begin{align}
\label{dist-h-decomp}
\mathbf{h}(k) &= \mathbf{h}_{\mathrm{D}}(k) + \overline{\mathbf{U}} (k) \ \overline{\mathbf{h}}(k),
\end{align}
where $\mathbf{h}_{\mathrm{D}}(k) = \mathbf{d}_{\theta{\mathrm{s}}}(k)/M$, and $\overline{\mathbf{U}}(k) \in \mathbb{C}^{M \times (M-1)}$ is a semi-unitary basis for the subspace orthogonal to $\mathbf{d}_{\theta{\mathrm{s}}}(k)$, i.e., $\overline{\mathbf{U}}^H(k) \overline{\mathbf{U}}(k)=\mathbf{I}_{M-1}$ and $\overline{\mathbf{U}}^H \mathbf{d}_{\theta_{\mathrm{s}}}(k)=\mathbf{0}$. The vector $\overline{\mathbf{h}} (k)\in \mathbb{C}^{(M-1)\times 1}$ collects the free parameters in this orthogonal subspace (e.g., obtained via a Gram–Schmidt construction).
Using this decomposition, the constrained problem can be reformulated as a quadratic eigenvalue problem whose solution leads to the following closed-form robust MVDR beamformer:
\begin{align}
	\label{S-beamf-2}
	\mathbf{h}_{\mathrm{RMVDR}} &= \mathbf{h}_{\mathrm{D}} - \overline{\mathbf{U}}
	\left( \overline{\mathbf{U}}^H \mathbf{\Gamma}_{\mathbf{v}}  \overline{\mathbf{U}}   - \lambda \mathbf{I}_{M-1} \right)^{-1}
	\overline{\mathbf{U}}^H \mathbf{\Gamma}_{\mathbf{v}} \mathbf{h}_{\mathrm{D}},
\end{align}
where $\lambda \in \mathbb{R}$ is uniquely determined by the WNG target $\mathcal{W}_0$ through the QEP (refer to~\cite{benesty2023microphone} for detail).

Consequently, the performance of robust MVDR beamforming critically depends on the accurate estimation of two key quantities: the noise spatial covariance matrix and the WNG threshold. While noise covariance estimation has been extensively studied, particularly through mask-based approaches, the selection of the WNG threshold (or equivalently, the diagonal loading factor) is often overlooked and typically determined empirically. An inappropriate choice of $\mathcal{W}_0$ can therefore lead to substantial performance degradation.
In practical systems, determining a suitable WNG threshold is especially challenging due to array inconsistencies, such as device-dependent microphone self-noise variations, which hinder the definition of a universal robustness setting across different arrays and acoustic conditions. This motivates the development of data-driven methods that can adaptively estimate the WNG constraint directly from the observed signals.

\section{Data-Driven Robustness Control for MVDR Beamforming}
\label{Sect-Method}

\subsection{Robust MVDR with Learnable WNG Constraints}
\label{Sect-Learnable-WNG}

To overcome the limitations of fixed robustness control and heuristic parameter tuning, a data-driven MVDR beamforming framework is proposed based on a dual-branch neural network architecture. The two branches are designed to address complementary mechanisms in robust beamforming: one branch estimates a frequency-dependent WNG constraint that directly controls robustness to array uncertainties, while the other branch predicts a complex-valued time--frequency (T--F) mask for noise covariance matrix estimation. By jointly learning these two quantities, the proposed framework enables adaptive robustness--directivity control and accurate spatial statistics estimation within a unified learning pipeline.

The network takes the short-time Fourier transform (STFT) coefficients of multi-channel speech signals as input. To extract informative representations suitable for both robustness control and covariance estimation, the feature extraction stage follows the multi-clue fusion principle proposed in~\cite{yang2023MCNET} and is implemented using the multi-channel JNF backbone introduced in~\cite{tesch22_interspeech}. Specifically, the feature extractor consists of four parallel modules that model T--F structures from complementary perspectives. The frequency module captures inter-frequency correlations, while the narrowband temporal module models short-term temporal dynamics along the time axis. The subband module exploits local frequency neighborhood expansion together with reference-channel information to characterize localized spectral patterns. In addition, the fullband module integrates cross-band information to capture long-term global context. The outputs of these modules are fused to form a unified multi-scale feature representation.

Each module adopts a unified RNN--FC architecture composed of a Bi-LSTM or LSTM layer followed by a fully connected (FC) layer and a ReLU activation. This design enforces structural consistency across modules while allowing each module to focus on distinct contextual cues. The resulting multi-scale features are shared by the two output branches, facilitating joint optimization and parameter efficiency.

The fused features are then fed into two task-specific prediction heads. The WNG branch employs a lightweight linear layer to predict a frequency-dependent robustness parameter, which specifies the desired WNG constraint for each frequency bin. In contrast, the complex mask branch uses a multilayer perceptron (MLP) to perform a nonlinear mapping from the shared features and estimates the real and imaginary components of the complex-valued T--F mask. This asymmetric design reflects the different functional roles of robustness control and spatial statistics estimation, while maintaining computational complexity.

Based on the output of the complex mask branch, the noise component 
$\widehat{\mathbf{v}}(k,l)$ is first estimated from the multichannel observation. 
The noise spatial covariance matrix is then computed by time averaging:
\begin{equation}
\widehat{\mathbf{\Phi}}_{\mathbf{v}}(k)
= \frac{1}{L} \sum_{l} 
\widehat{\mathbf{v}}(k,l)\widehat{\mathbf{v}}^{\mathrm{H}}(k,l),
\label{eq:scm_n}
\end{equation}
where $L$ denotes the number of time frames. 
Since each outer product 
$\widehat{\mathbf{v}}(k,l)\widehat{\mathbf{v}}^{\mathrm{H}}(k,l)$
is Hermitian positive semi-definite, the averaged covariance matrix 
$\widehat{\mathbf{\Phi}}_{\mathbf{v}}(k)$ also remains Hermitian positive semi-definite.Therefore, it forms a valid noise covariance estimate for the subsequent MVDR beamformer design.

\subsection{Training with Differentiable Robust MVDR}
\label{Sect-Training}

The proposed framework is trained in an end-to-end manner by embedding a differentiable WNG-constrained MVDR beamforming layer into the learning pipeline. The training objective is defined as the mean absolute error (MAE) between the enhanced output signal and an early-reference beamformed signal:
\begin{equation}
\mathcal{L}_{\text{total}}
= \frac{1}{N} \sum_{i=1}^{N}
\left| y_{\text{early}}^{(i)} - y_{\text{filtered}}^{(i)} \right|_1,
\end{equation}
where $y_{\text{filtered}}^{(i)}$ denotes the output of the proposed differentiable robust MVDR layer for the $i$-th training sample, $y_{\text{early}}^{(i)}$ represents the corresponding early-reference signal, and $N$ is the total number of training samples.

Although it is impractical to provide explicit supervision for the WNG by manually specifying a target value for each training utterance, the predicted WNG is implicitly constrained through the differentiable beamforming operation and the reconstruction loss. The WNG governs the trade-off between robustness and spatial selectivity: excessively large values increase robustness at the expense of reduced directivity, whereas overly small values increase sensitivity to array mismatches and microphone self-noise. During training, the predicted WNG directly affects the beamformer output, which is compared against the early-reference signal. This mechanism naturally guides the network toward physically meaningful robustness levels, enabling stable and interpretable data-driven control that adapts to the acoustic scene and array characteristics.

\section{Experimental Results}

\begin{figure}[htbp]
  \centering
  \includegraphics[width=0.45\textwidth]{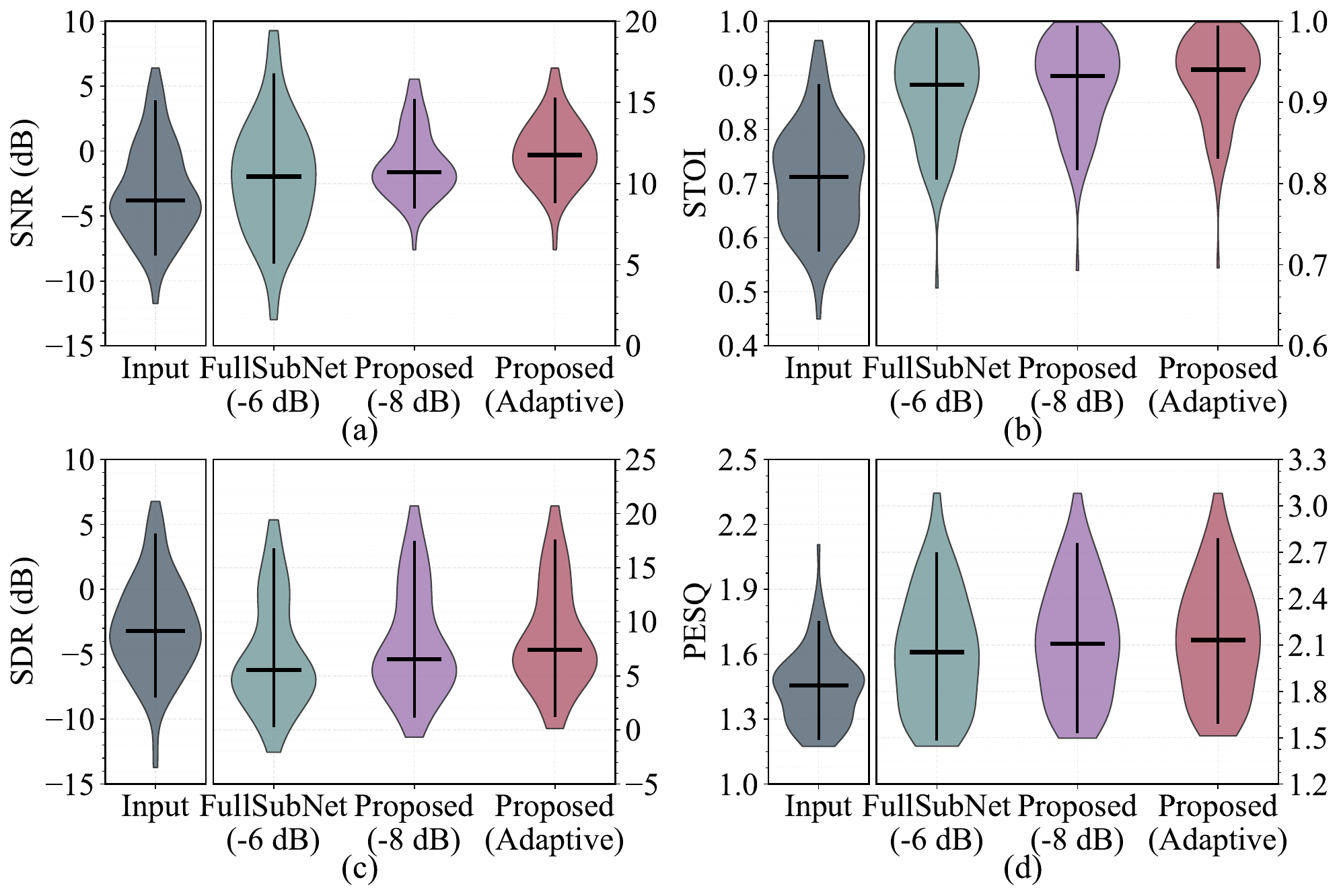}
  \caption{Comparison of objective metrics: (a) SNR, (b) STOI, (c) SDR, and (d) PESQ. Violin plots show the distribution of utterance-level scores for the input signal, FullSubNet with its optimal WNG setting (-6 dB), and the proposed methods using the optimal fixed WNG (-8 dB) and the adaptive WNG strategy.}
  \label{fig:violinplot}
\end{figure}

\subsection{Dataset and Acoustic Experimental Setup}

The VCTK dataset is used as the speech source, which is sampled at $16$kHz and from multiple speakers. Each target speech segment is truncated to a fixed duration of $3$s. To generate multichannel noisy signals, an $8$-microphone ULA with an inter-microphone spacing of $2$cm is employed. The target source is positioned in the endfire direction. 
Room dimensions are randomly sampled with lengths in $[5,10]$ m, widths in $[4,8]$ m, and heights in $[2.5,4]$ m. The reverberation time $T_{60}$ is uniformly drawn from $0.1$ to $0.4$s.
The number of interfering sources is randomly selected between $1$ and $4$, with azimuths uniformly distributed from $90^\circ$ to $270^\circ$. The signal-to-interference ratio (SIR) is randomly chosen from $0$ to $10$dB. 
In addition, spatially diffuse noise and additive white Gaussian noise are added, with SNRs randomly selected between $0–10$dB and $10–40$dB, respectively. Both SNRs are defined with respect to the target signal power. Target speech and interfering sources fully overlap in time and are convolved with room impulse responses under varying acoustic conditions to produce multichannel reverberant signals.
All time-domain signals are transformed into the STFT domain using a frame length of $16$ms with $50\%$ overlap. The FFT length is set to $256$. 

Network hyperparameters follow the configuration in~\cite{yang2023MCNET}. The LSTM layers in the four modules contain $128$, $256$, $384$, and $128$ hidden units, respectively. The third module uses $N_1=2$ adjacent frequency bins, and the fourth module uses $N_2=5$ contextual frames.
The model is trained using the Adam optimizer~\cite{kingma2014adam} at the utterance level with a batch size of $4$. The initial learning rate is $10^{-4}$ and is halved when the validation loss does not improve for $5$ consecutive epochs. 
Performance is evaluated using four objective metrics: SNR gain, signal-to-distortion ratio (SDR)~\cite{Vincent2006Performance}, short-time objective intelligibility (STOI)~\cite{STOI2010}, and perceptual evaluation of speech quality (PESQ)~\cite{rix2001perceptual}.

The first experiment compares the SNR, STOI, SDR, and PESQ performance of the conventional and proposed MVDR beamformers.
For the conventional MVDR beamformers, the time--frequency mask is estimated using a FullSubNet model trained on the
VCTK dataset~\cite{hao2021fullsubnet}, which is also used to train the proposed network. The model takes
the noisy signal from a single reference microphone channel as input and predicts a time--frequency mask that is shared across all channels. The estimated mask is then used to compute the noise covariance matrix according to~\eqref{eq:scm_n}.We report the best-performing fixed WNG setting for this baseline, which is achieved at $\mathcal{W}_0=-6$ dB. 
For the proposed MVDR beamformers, the time–frequency mask and $\mathcal{W}_0$ are estimated using the proposed model, and the noise covariance matrix is computed following the procedure as in the conventional case. We report results with the adaptive WNG strategy and with the best fixed WNG setting ($\mathcal{W}_0=-8$ dB).
Figure~\ref{fig:violinplot} shows plots of the results. Compared with the conventional baseline under its best fixed WNG setting, the proposed method yields improved performance across all metrics, while the adaptive WNG strategy provides more robust results.

\begin{table}[t]
    \caption{Performance comparison of MVDR-based methods under both seen and unseen array conditions.
    SNR gain and $\Delta$SDR are measured in dB.}
    \label{tab:mvdr_dl_performance_mismatch}
    \centering
    \small
    \setlength{\tabcolsep}{4pt}
    \renewcommand{\arraystretch}{1.15}
    
    \begin{tabular*}{\columnwidth}{@{\extracolsep{\fill}}lcc}
        \toprule
        \textbf{Configuration} & \textbf{SNR gain} & \boldmath$\Delta$\textbf{SDR} \\
        \midrule
        
        \multicolumn{3}{l}{\textbf{Seen array conditions} ($\delta = 2.0 \pm \epsilon$~cm)} \\
        \midrule
        Proposed MVDR & 11.940 & 11.474 \\
        Conventional MVDR (with optimal $\varepsilon$)  & 10.118 & 9.275 \\
        Conventional MVDR (with optimal $\mathcal{W}_0$)   & 10.543 & 9.510 \\
        \midrule
        \multicolumn{3}{l}{\textbf{Unseen array conditions} ($\delta = 1.0 \pm \epsilon$~cm)} \\
        \midrule
        Proposed MVDR & 10.225 & 9.93 \\
        Conventional MVDR (with optimal $\varepsilon$) 
        & 8.883 & 8.701 \\
        Conventional MVDR (with optimal $\mathcal{W}_0$)  & 8.683 & 8.476 \\
        \midrule
        \multicolumn{3}{l}{\textbf{Unseen array conditions} ($\delta = 3.0 \pm \epsilon$~cm)} \\
        \midrule
        Proposed MVDR & 11.586 & 10.850 \\
        Conventional MVDR (with optimal $\varepsilon$) 
        & 9.889 & 8.649 \\
        Conventional MVDR (with optimal $\mathcal{W}_0$)  & 9.952 & 8.786 \\
        \bottomrule
    \end{tabular*}
\end{table}

In the second experiment, random array mismatch is introduced to assess robustness. The seen array configuration adopts a nominal inter-element spacing of $2.0$ cm, while the unseen array conditions use nominal spacings of $1.0$ cm and $3.0$ cm. To simulate practical array perturbations, each configuration is modeled as $\delta = d + \epsilon$, where $\epsilon$ follows a zero-mean Gaussian distribution with a standard deviation of $0.1$ cm. The corresponding results are reported in Table~\ref{tab:mvdr_dl_performance_mismatch}.
For the conventional methods, the optimal WNG constraint (or equivalently, the optimal diagonal loading factor) is manually tuned. As shown in Table~\ref{tab:mvdr_dl_performance_mismatch}, even under their optimal settings, the conventional approaches underperform the proposed method. 
These results demonstrate that the proposed framework can automatically adapt to varying noise conditions and array mismatches by jointly improving covariance estimation and WNG control, leading to more stable and enhanced speech enhancement performance.

\section{Conclusion}

\label{Sec-Con}
This work proposed a data-driven method for estimating the WNG constraint in MVDR beamforming. Unlike conventional approaches that use a fixed WNG threshold, the proposed framework employs a deep neural network to jointly predict the optimal WNG value and the noise presence mask. By doing so, the beamformer can dynamically adjust its robustness to microphone mismatch while maintaining directivity for noise and interference suppression according to the prevailing acoustic conditions. Extensive experiments demonstrate that the proposed approach consistently outperforms fixed-threshold baselines across a range of noisy and reverberant scenarios. These results indicate that data-driven WNG estimation is a promising direction for improving the adaptability and effectiveness of MVDR beamformers in real-world applications.

\section{Acknowledgments}
This work was supported by the National Natural Science Foundation (NSFC) of China under Grant 62471340. The numerical calculations in this paper have been done on the supercomputing system in the Supercomputing Center of Wuhan University.

\section{Generative AI Use Disclosure}

ChatGPT was used only for language polishing and grammar checking.

\bibliographystyle{IEEEtran}
\bibliography{Bib_MABFSE}

\end{document}